\documentclass[twocolumn]{llncs}

\usepackage[left=2cm, right=2cm, top=3cm, bottom=3cm]{geometry}

\usepackage[disable]{todonotes}
\usepackage[inline]{enumitem}
\usepackage[abbreviations, hyperfirst=false]{glossaries-extra}
\usepackage{todonotes}
\usepackage{xspace}
\usepackage{booktabs}
\usepackage{subfig}
\usepackage{balance}

\usepackage{textcomp}

\usepackage{tikz}
\usetikzlibrary{calc,arrows.meta,fit,shapes.geometric,shapes.arrows}
\tikzstyle{container}=[draw,rounded corners,fill=blue!20,align=center,thick]
\tikzstyle{volume}=[draw,rounded corners,fill=yellow!20,align=center,thick]
\tikzstyle{container-fixed}=[draw,rounded corners,fill=blue!20,align=center,thick,minimum width=2cm,minimum height=2cm]
\tikzstyle{volume-fixed}=[draw,rounded corners,fill=yellow!20,align=center,thick,minimum width=2cm,minimum height=2cm]
\tikzstyle{tc-1}=[draw,rounded corners,fill=red!20,align=center,thick]
\tikzstyle{simulation}=[draw,rounded corners,fill=yellow!20,align=center,thick]
\tikzstyle{tc-2}=[draw,rounded corners,fill=green!20,align=center,thick]
\tikzstyle{db}=[draw,cylinder,shape border rotate=90,minimum height=.75cm,minimum width=.5cm,fill=white]
\tikzstyle{ros-bridge}=[draw,fill=green!20]
\tikzstyle{program}=[font={\usebox{\programbox}},inner sep = 0]
\tikzstyle{programs}=[font={\usebox{\programsbox}},inner sep = 0]
\tikzstyle{monitor}=[font={\usebox{\lookingglassbox}},inner sep=0]
\tikzstyle{simulator}=[font={\usebox{\simulatorbox}},inner sep=0]
\tikzstyle{simulators}=[font={\usebox{\simulatorsbox}},inner sep=0]
\tikzstyle{document}=[font={\usebox{\documentbox}},inner sep=0]
\tikzstyle{documents}=[font={\usebox{\documentsbox}},inner sep=0]
\tikzstyle{videos}=[font={\usebox{\videosbox}},inner sep = 0]
\tikzstyle{user}=[font={\usebox{\userbox}},inner sep = 0]
\tikzstyle{llm}=[font={\usebox{\llmbox}},inner sep = 0]
\tikzstyle{screenshot}=[font={\usebox{\screenshotbox}},inner sep = 0]

\pgfdeclarelayer{background}
\pgfdeclarelayer{foreground}
\pgfsetlayers{background,main,foreground}

\newsavebox{\userbox}
\savebox\userbox{
\begin{tikzpicture}[outer sep=0,inner sep=0,thick]
  \draw (0,0) circle (.15);
  \draw (0,-.15) -- (0,-.5);
  \draw (0,-.5) -- (-.2,-.7);
  \draw (0,-.5) -- (.2,-.7);
  \draw (0,-.35) -- (-.2,-.15);
  \draw (0,-.35) -- (.2,-.15);
\end{tikzpicture}
}

\newsavebox{\screenshotbox}
\savebox\screenshotbox{
\begin{tikzpicture}[outer sep=0,inner sep=0,scale=.75]
  \draw[thick,rounded corners=.1cm,line width=.05cm,fill=white]
    (0,0) rectangle (1.5,1);
  \draw[thick,rounded corners=.1cm,line width=.05cm,fill=white]
    (.1,.1) rectangle (1.4,.9);
  \draw[thick,line width=.05cm,rounded corners=.1cm]
    (.1,.4) -- (.3,.8) -- (.7,.2) -- (1.1,.5) -- (1.4,.3);
  \draw[thick]
    (1.2, .7) circle (.1);
  \draw[thick] (-.1,.9) |- (.1,1.1);
  \draw[thick] (-.1,.1) |- (.1,-.1);
  \draw[thick] (1.4,1.1) -| (1.6,.9);
  \draw[thick] (1.4,-.1) -| (1.6,.1);
\end{tikzpicture}
}

\newsavebox{\desktopbox}
\savebox\desktopbox{
\begin{tikzpicture}[outer sep=0,inner sep=0]
  \draw[thick,rounded corners=.1cm,line width=.05cm,fill=white]
    (0,0) -- (.5,0) -- (.5,.7) -- (-.5,.7) -- (-.5,0) -- cycle;
    
  \draw[thick,rounded corners] (0,0) -- (0,-.2);
  \draw[thick,rounded corners] (-.3,-.18) -- (.3,-.18);
\end{tikzpicture}
}

\newsavebox{\documentbox}
\savebox{\documentbox}{\begin{tikzpicture}[scale=1.75,thick]
	\draw[thick,fill=white] (0,0) -- (0,-.5) -- (.4,-.5) -- (.4,-.1) -- (.3,0) -- cycle;
	\draw[thick] (.3,0) -- (.3,-.1) -- (.4,-.1);
	\foreach \y in {-.1,-.15,...,-.4} {
		\draw[thin] (.05,\y) -- (.35,\y);
	}
\end{tikzpicture}}

\newsavebox{\documentsbox}
\savebox{\documentsbox}{\begin{tikzpicture}[scale=2,thick]
    \begin{scope}[xshift=-.05cm,yshift=.05cm]
      \draw[thick,fill=white] (0,0) -- (0,-.5) -- (.4,-.5) -- (.4,-.1) -- (.3,0) -- cycle;
      \draw[thick] (.3,0) -- (.3,-.1) -- (.4,-.1);
      \foreach \y in {-.1,-.15,...,-.4} {
        \draw[thin] (.05,\y) -- (.35,\y);
      }
      \draw[white,fill=white,fill opacity=.4,draw opacity=.4] (0,0) -- (0,-.5) -- (.4,-.5) -- (.4,-.1) -- (.3,0) -- cycle;
      \draw[white,fill=white,fill opacity=.4,draw opacity=.4] (0,0) -- (0,-.5) -- (.4,-.5) -- (.4,-.1) -- (.3,0) -- cycle;
    \end{scope}

    \begin{scope}
      \draw[thick,fill=white] (0,0) -- (0,-.5) -- (.4,-.5) -- (.4,-.1) -- (.3,0) -- cycle;
      \draw[thick] (.3,0) -- (.3,-.1) -- (.4,-.1);
      \foreach \y in {-.1,-.15,...,-.4} {
        \draw[thin] (.05,\y) -- (.35,\y);
      }
      \draw[white,fill=white,fill opacity=.4,draw opacity=.4] (0,0) -- (0,-.5) -- (.4,-.5) -- (.4,-.1) -- (.3,0) -- cycle;
    \end{scope}

    \begin{scope}[xshift=.05cm,yshift=-.05cm]
      \draw[thick,fill=white] (0,0) -- (0,-.5) -- (.4,-.5) -- (.4,-.1) -- (.3,0) -- cycle;
      \draw[thick] (.3,0) -- (.3,-.1) -- (.4,-.1);
      \foreach \y in {-.1,-.15,...,-.4} {
        \draw[thin] (.05,\y) -- (.35,\y);
      }
    \end{scope}
\end{tikzpicture}}

\newsavebox{\workstationbox}
\savebox\workstationbox{
\begin{tikzpicture}[thick]
  \draw (0,0) rectangle (.75,1);

  \draw (.1,.9) rectangle (.65,.8);
  \draw (.1,.8) rectangle (.65,.7);
  \draw (.1,.7) rectangle (.65,.6);

  \draw (.375,.2) circle (.08);
\end{tikzpicture}
}

\newsavebox{\serverbox}
\savebox\serverbox{
\begin{tikzpicture}[thick]
	\node[draw,rounded corners=.1cm,minimum height=.075cm,minimum width=1.2cm,fill=white] at (0,0) {};
	\node[circle,minimum size=.1cm,fill=black,inner sep=0] at (-.4cm,0cm) {};
	
	\node[draw,rounded corners=.1cm,minimum height=.075cm,minimum width=1.2cm,fill=white] at (0,.35) {};
	\node[circle,minimum size=.1cm,fill=black,inner sep=0] at (-.4cm,.35) {};
	
	\node[draw,rounded corners=.1cm,minimum height=.075cm,minimum width=1.2cm,fill=white] at (0,.7) {};
	\node[circle,minimum size=.1cm,fill=black,inner sep=0] at (-.4cm,.7) {};
\end{tikzpicture}
}

\newsavebox{\quantumbox}
\savebox\quantumbox{
\begin{tikzpicture}[outer sep=0,inner sep=0,scale=2]
  \draw[thick,fill=white] (0,0) circle (.05cm);
  \draw[thick] (0,0) ellipse (.2cm and .08cm);
  \draw[thick,rotate=60] (0,0) ellipse (.2cm and .08cm);
  \draw[thick,rotate=-60] (0,0) ellipse (.2cm and .08cm);
\end{tikzpicture}
}

\newsavebox{\circuitbox}
\savebox\circuitbox{
\begin{tikzpicture}[outer sep=0,inner sep=0,scale=.5]
  \draw (0,0) -- (1.5,0);
  \draw (0,-1) -- (1.5,-1);

  \draw[fill=black] (.75,0) circle (.1);
  \draw[] (.75,-1) circle (.2);
  \draw (.75,-1.2) -- (.75,0);

\end{tikzpicture}
}

\newsavebox{\programbox}
\savebox\programbox{\begin{tikzpicture}[outer sep=0,inner sep=0,thick,scale=.75]
  \draw[rounded corners,fill=white] (1,0) -| (1.5,-1) -| (0,0) -- (1,0);
  \node[inner sep=0,outer sep=0] at (.75,-.5) {$>$\_};
\end{tikzpicture}}

\newsavebox{\programsbox}
\savebox\programsbox{
\begin{tikzpicture}[outer sep=0,inner sep=0,thick]
  \begin{scope}[xshift=-.2cm,yshift=.2cm]
    \draw[rounded corners] (1,0) -| (1.5,-1) -| (0,0) -- (1,0);
    \node at (.75,-.5) {$>$\_};

    \draw[fill=white,draw=white, fill opacity = .4, draw opacity = .4, rounded corners] (1,0) -| (1.5,-1) -| (0,0) -- (1,0);
    \draw[fill=white,draw=white, fill opacity = .4, draw opacity = .4, rounded corners] (1,0) -| (1.5,-1) -| (0,0) -- (1,0);
  \end{scope}

  \begin{scope}
    \draw[rounded corners,fill=white] (1,0) -| (1.5,-1) -| (0,0) -- (1,0);
    \node at (.75,-.5) {$>$\_};
    \draw[fill=white,draw=white, fill opacity = .4, draw opacity = .4, rounded corners] (1,0) -| (1.5,-1) -| (0,0) -- (1,0);
  \end{scope}

  \begin{scope}[xshift=.2cm,yshift=-.2cm]
    \draw[rounded corners,fill=white] (1,0) -| (1.5,-1) -| (0,0) -- (1,0);
    \node at (.75,-.5) {$>$\_};
  \end{scope}

\end{tikzpicture}
}

\newsavebox{\browserbox}
\savebox\browserbox{
\begin{tikzpicture}[outer sep=0,inner sep=0]
  \draw[rounded corners] (1,0) -| (1.5,-1) -| (0,0) -- (1,0);
  \draw (0,-.1) -- (1.5,-.1);

  \draw (.75,-.55) circle (.3cm);
  \draw (.45,-.55) -- (1.05,-.55);
  \draw (.75,-.55) ellipse (.15cm and .3cm);
\end{tikzpicture}
}

\newcommand{\gear}[6]{%
  (0:#2)
  \foreach \i [evaluate=\i as \n using {\i-1)*360/#1}] in {1,...,#1}{%
    arc (\n:\n+#4:#2) {[rounded corners=1pt] -- (\n+#4+#5:#3)
    arc (\n+#4+#5:\n+360/#1-#5:#3)} --  (\n+360/#1:#2)
  } -- cycle %
}

\newsavebox{\gearbox}
\savebox{\gearbox}{\begin{tikzpicture}
\begin{scope}[shift={(-.2cm, .2cm)}]
\draw[thick,fill=white] \gear{8}{.4cm}{.5cm}{22.5}{1}{0};
\end{scope}
\draw[thick,fill=white] \gear{8}{.4cm}{.5cm}{22.5}{1}{0};
\end{tikzpicture}}

\newsavebox{\lookingglassbox}
\savebox{\lookingglassbox}{\begin{tikzpicture}[outer sep=0, inner sep=0,thick]
  \draw (0,0) -- ++(-.25,-.25) arc (135:315:.0625) -- ++(.25,.25);
  \draw[fill=white] (.125,.0625) circle (.25);
  \draw[fill=white] (.125,.0625) circle (.2);
\end{tikzpicture}}

\newsavebox{\videosbox}
\savebox{\videosbox}{ \begin{tikzpicture}[outer sep=0,inner sep=0,thick]
  \begin{scope}[xshift=-.2cm,yshift=.2cm]

  \end{scope}

  \begin{scope}
    \draw[rounded corners,fill=white] (1,0) -| (1.5,-1) -| (0,0) -- (1,0);
    \draw (.2,0) -- (.2,-1);
    \draw (1.3,0) -- (1.3,-1);

    \draw (.1,-.2) circle (.05);
    \draw (.1,-.4) circle (.05);
    \draw (.1,-.6) circle (.05);
    \draw (.1,-.8) circle (.05);

    \draw (1.4,-.2) circle (.05);
    \draw (1.4,-.4) circle (.05);
    \draw (1.4,-.6) circle (.05);
    \draw (1.4,-.8) circle (.05);

    \draw[draw=white,fill=white,draw opacity=.4, fill opacity=.4] (1,0) -| (1.5,-1) -| (0,0) -- (1,0);
  \end{scope}

  \begin{scope}[xshift=.2cm,yshift=-.2cm]
    \draw[rounded corners,fill=white] (1,0) -| (1.5,-1) -| (0,0) -- (1,0);
    \draw (.2,0) -- (.2,-1);
    \draw (1.3,0) -- (1.3,-1);

    \draw (.1,-.2) circle (.05);
    \draw (.1,-.4) circle (.05);
    \draw (.1,-.6) circle (.05);
    \draw (.1,-.8) circle (.05);

    \draw (1.4,-.2) circle (.05);
    \draw (1.4,-.4) circle (.05);
    \draw (1.4,-.6) circle (.05);
    \draw (1.4,-.8) circle (.05);
  \end{scope}
\end{tikzpicture}}

\newsavebox{\loadbalancerbox}
\savebox\loadbalancerbox{\begin{tikzpicture}[outer sep=0,inner sep=0,thick]
    \node[fill=black,minimum size=.1cm] (1) at (0,0) {};
    \node[fill=black,minimum size=.1cm] (2) at (-1/4,-1/2) {};
    \node[fill=black,minimum size=.1cm] (3) at (0,-1/2) {};
    \node[fill=black,minimum size=.1cm] (4) at (1/4,-1/2) {};

    \path[draw,thick,-,rounded corners] (1) -- (0,-1/4) -| (2);
    \path[draw,thick,-] (1) -- (0,-1/4) -| (3);
    \path[draw,thick,-,rounded corners] (1) -- (0,-1/4) -| (4);
\end{tikzpicture}}

\newsavebox{\llmbox}
\savebox\llmbox{\begin{tikzpicture}[outer sep=0,inner sep=0,thick,scale=.25]
    \node[draw,circle] (0-0) at (0,.5) {};
    \node[draw,circle] (0-1) at (0,1.5) {};

    \node[draw,circle] (1-0) at (1,0) {};
    \node[draw,circle] (1-1) at (1,1) {};
    \node[draw,circle] (1-2) at (1,2) {};

    \node[draw,circle] (2-0) at (2,1) {};

    \foreach \x in {0,1} {
      \foreach \y in {0,1,2} {
        \draw[-] (0-\x) -- (1-\y);
      }
    }

    \foreach \x in {0,1,2} {
      \draw[-] (1-\x) -- (2-0);
    }
\end{tikzpicture}}

\newsavebox{\simulatorbox}
\savebox\simulatorbox{\begin{tikzpicture}[outer sep=0,inner sep=0,thick]
  \draw[rounded corners,fill=white] (1,0) -| (1.5,-1) -| (0,0) -- (1,0);
  \draw (0,-1/4) -- (1.5,-1/4);
  \draw[dashed] (0,-1/2) -- (1.5,-1/2);
  \draw (0,-3/4) -- (1.5,-3/4);
\end{tikzpicture}}

\newsavebox{\simulatorsbox}
\savebox\simulatorsbox{\begin{tikzpicture}[outer sep=0,inner sep=0,thick]
  \begin{scope}[xshift=-.2cm,yshift=.2cm]
    \draw[rounded corners,fill=white] (1,0) -| (1.5,-1) -| (0,0) -- (1,0);
    \draw (0,-1/4) -- (1.5,-1/4);
    \draw[dashed] (0,-1/2) -- (1.5,-1/2);
    \draw (0,-3/4) -- (1.5,-3/4);
    \draw[rounded corners,fill=white,draw=white,draw opacity=.4,fill opacity=.4] (1,0) -| (1.5,-1) -| (0,0) -- (1,0);
    \draw[rounded corners,fill=white,draw=white,draw opacity=.4,fill opacity=.4] (1,0) -| (1.5,-1) -| (0,0) -- (1,0);
  \end{scope}

  \begin{scope}
    \draw[rounded corners,fill=white] (1,0) -| (1.5,-1) -| (0,0) -- (1,0);
    \draw (0,-1/4) -- (1.5,-1/4);
    \draw[dashed] (0,-1/2) -- (1.5,-1/2);
    \draw (0,-3/4) -- (1.5,-3/4);
    \draw[rounded corners,fill=white,draw=white,draw opacity=.4,fill opacity=.4] (1,0) -| (1.5,-1) -| (0,0) -- (1,0);
  \end{scope}

  \begin{scope}[xshift=.2cm,yshift=-.2cm]
    \draw[rounded corners,fill=white] (1,0) -| (1.5,-1) -| (0,0) -- (1,0);
    \draw (0,-1/4) -- (1.5,-1/4);
    \draw[dashed] (0,-1/2) -- (1.5,-1/2);
    \draw (0,-3/4) -- (1.5,-3/4);
  \end{scope}
\end{tikzpicture}}

\newcommand\createabbrv[3]{%
    \newabbreviation{#1}{#2}{#3}%
    \expandafter\newcommand\csname #2\endcsname{\gls{#1}\xspace}%
    \expandafter\newcommand\csname #2s\endcsname{\glspl{#1}\xspace}%
}

\newcommand{\RCE}{RCE\xspace}
\newcommand{\JSON}{JSON\xspace}
\createabbrv{dlr}{DLR}{German Aerospace Center}
\createabbrv{llm}{LLM}{Large Language Model}
\createabbrv{gui}{GUI}{Graphical User Interface}
\createabbrv{ocr}{OCR}{Optical Character Recognition}
\createabbrv{rcp}{RCP}{Eclipse Rich Client Platform}
\createabbrv{os}{OS}{Operating System}
\createabbrv{ai}{AI}{Artificial Intelligence}

\newcommand{\etal}{et al.\xspace}
\newcommand{\GPTDroid}{GPT Droid\xspace}
\newcommand{\GERALLT}{\textsc{GERALLT}\xspace}

\title{Automated Testing of the GUI of a Real-Life Engineering Software using Large Language Models\thanks{This work is based on the bachelor thesis of the first author.~\cite{Rosenbach2024}}}

\author{Tim Rosenbach\inst{1}\and David Heidrich\inst{2}\and Alexander Weinert\inst{1}}

\institute{
German Aerospace Center (DLR), Institute of Software Technology, Cologne, Germany\\
\email{\{tim.rosenbach, alexander.weinert\}@dlr.de}
\and
German Aerospace Center (DLR), Institute of Software Technology, Oberpfaffenhofen, Germany\\
\email{david.heidrich@dlr.de}
}

\begin{document}

\maketitle

\begin{abstract}
One important step in software development is testing the finished product with actual users.
These tests aim, among other goals, at determining unintuitive behavior of the software as it is presented to the end-user.
Moreover, they aim to determine inconsistencies in the user-facing interface.
They provide valuable feedback for the development of the software, but are time-intensive to conduct.

In this work, we present \GERALLT, a system that uses \LLMs to perform exploratory tests of the \GUI of a real-life engineering software.
\GERALLT automatically generates a list of potential unintuitive and inconsistent parts of the interface.
We present the architecture of \GERALLT and evaluate it on a real-world use case of the engineering software, which has been extensively tested by developers and users.
Our results show that \GERALLT is able to determine issues with the interface that support the software development team in future development of the software.
\end{abstract}

\keywords{Software Testing \and Graphical User Interface \and Acceptance Testing \and Large Language Models}

\glsreset{llm}
\glsreset{gui}

\section{Introduction}

When developing software, large efforts are spent on quality assurance~\cite{GarousiZhi2013}, which is mainly performed by conducting a number of automated and manual tests on the software.
Developers use such tests to determine the compliance of the software with domain requirements as well as to determine the quality of the software.
Popular qualitative metrics for the software under development are, e.g., its performance, security, or accessibility~\cite{ISO2023}.
To varying degrees, such tests can be automated and be executed and evaluated without human input.

One major aspect of software quality, which is not easily captured via automated tests, is the acceptance of the software by end users.
User acceptance depends not only on the satisfaction of domain requirements, but also, among others, on the intuitiveness and consistency of the user interface.
These metrics are hard to capture using automated tests.
Hence, these criteria are often tested for using manual tests of the interface of the software.
While manual tests are perceived as effective by software developers, they are also expensive in terms of time and resources spent~\cite{HaasElsnerJuergensEtAl2021}.
Although some work exists on the automation of such tests, that work is mainly focused on either web-based software, or applications for popular mobile operating systems, like Android.~\cite{LiuChenWangEtAl2024,ZimmermannKoziolek2023,BergsmannSchmidtFischerEtAl2024,ZimmermannDeubelKoziolek2023,ZhangZhengBaiEtAl2024,GuanBaiLiu2024,XueLiTianEtAl2024}

In this work, we present \GERALLT, a system that aims to support developers in testing the interface of a real-life desktop-based engineering software that is in productive use throughout numerous institutes of the \DLR. 
At the core of \GERALLT are two components based on \LLMs, a type of generative \AI capable of producing human-like text based on patterns learned from vast amounts of data.
One of these components explores the \GUI, while the other evaluates the observed behavior for unintuitive or inconsistent interactions, as well as for functional errors.
We then evaluate the issues determined by \GERALLT via discussion with the developers of the software.

The main contribution of this work consists of an architecture description of \GERALLT and its evaluation.
\GERALLT comprises multiple components and is partially based on previous work due to Lui et al.~\cite{LiuChenWangEtAl2024}.
We evaluate \GERALLT by testing a specific feature of the engineering software with it and discussing the issues \GERALLT determined with the software engineers.

The remainder of this work is organized as follows:
After giving an overview over previous work in Section~\ref{sec:related-work} we describe the system under test, its use at \DLR, and its technical implementation in Section~\ref{sec:system}.
Afterwards, we evaluate \GERALLT in Section~\ref{sec:evaluation} and conclude this work with a summary of our results and perspectives for future work in Section~\ref{sec:conclusion}.

\section{Background and Related Work}
\label{sec:related-work}

\LLMs are a class of generative \AI, which are trained on large amounts of textual data and refined through human feedback.
Given a textual input prompt, they produce textual responses, which mimic human responses with surprising accuracy in a diverse array of tasks.
In recent years, there has been a flurry of research on new domains in which \LLMs can be applied.~\cite{KaurKashyapKumarEtAl2024}
In particular, there exist a number of works on the application of \LLMs in software quality assurance.
Among others, these include using \LLMs for generating~\cite{GarciaLeottaRiccaEtAl2024,GuanBaiLiu2024,XueLiTianEtAl2024} and executing~\cite{BergsmannSchmidtFischerEtAl2024,TianShuWangEtAl2024} test cases, for exploring the system under test, e.g., via fuzzing~\cite{ZhangZhengBaiEtAl2024}, and for repairing detected issues~\cite{XiaZhang2024,YangTianPianEtAl2024,ShanHuoSuEtAl2024}.

Moreover, there exist numerous works on the automation of \GUI testing~\cite{FulciniGaraccioneCoppolaEtAl2022}.
Of particular interest are novel strategies for the traversal of applications for the Android operating system~\cite{Wong2023,GuRojas2023} as well as the use of \AI for the traversal of arbitrary \GUIs~\cite{ZimmermannDeubelKoziolek2023,ZimmermannKoziolek2023}.
In the remainder of this section, we focus on three pieces of related work that align most closely with our work presented here.

Liu \etal~\cite{LiuChenWangEtAl2024} developed \GPTDroid, an \LLM-based system for testing Android apps.
\GPTDroid extracts the \GUI context from an app and encodes it into prompts for the \LLM.
The \LLM generates operational commands, which \GPTDroid translates into executable actions.
\GPTDroid includes a functionality-aware memory mechanism to retain long-term knowledge of the testing process.
This mechanism allows the \LLM to guide exploration based on app functionality.
In evaluations on 93 Android apps, \GPTDroid achieved 75\% activity coverage and detected 31\% more bugs than the best baseline.
It also identified 53 new bugs on Google Play, with developers confirming or fixing 35 of them.

Gao \etal~\cite{GaoJiBaiEtAl2023} developed a system for automating desktop \GUI tasks by utilizing \LLMs.
The system includes a \GUI Parser that converts screenshots and metadata into a structured representation by combining \OCR, icon detection, and other vision tools.
This representation enables the system to understand diverse UI elements and their spatial relationships.
Their experiments showed the framework achieving a 46\% success rate, revealing the challenges and potential for improvements in this domain.

Zimmermann and Koziolek~\cite{ZimmermannKoziolek2023} use \LLMs to test a web-based example application.
Web-based applications have a natural textual representation, which is rendered by the browser.
In contrast, we investigate a desktop-based application where we first need to generate a textual representation of the current state of the \GUI.
Moreover, here we investigate the capabilities of \LLMs for testing a real-world application instead of an example application specifically constructed for the purpose of testing.

Our work builds upon the approaches of Liu \etal and Gao \etal.
We use the concept of Liu \etal for automated \GUI testing and combine it with the approach of Gao \etal for the automation of a desktop application.
The key novelty of our work lies in automating \GUI testing for a desktop application.
While prior research has focused on mobile and web applications, our approach extends \LLM-based testing techniques to a real-world desktop environment.

\section{The Use Case}
\label{sec:use-case}

\RCE is an open-source software used for designing, implementing, and executing simulation toolchains.
Engineers at \DLR use \RCE as part of their daily work to simulate complex systems, such as aircraft, ships, and satellites~\cite{BodenFlinkFoerstEtAl2021,FlinkMischkeSchaffertEtAl2022}.
\RCE itself does not provide any discipline-specific simulations ``out of the box'' but instead relies on engineers \emph{integrating} their pre-existing simulations into \RCE.
Integrating a simulation into \RCE mainly amounts to defining inputs and outputs of the simulation as well as providing a shell script defining the invocation of the simulation.
While straightforward in concept, there are numerous additional decisions to be made by the tool integrator, such as defining correct paths for execution, setting up the environment of the simulation, or allowing for concurrent executions of simulations.
In practice, the integration of an existing simulation is a complex task that requires in-depth knowledge of both \RCE and the simulation to be integrated.

\begin{figure*}
  \subfloat[First Page]{
    \includegraphics[width=.32\linewidth]{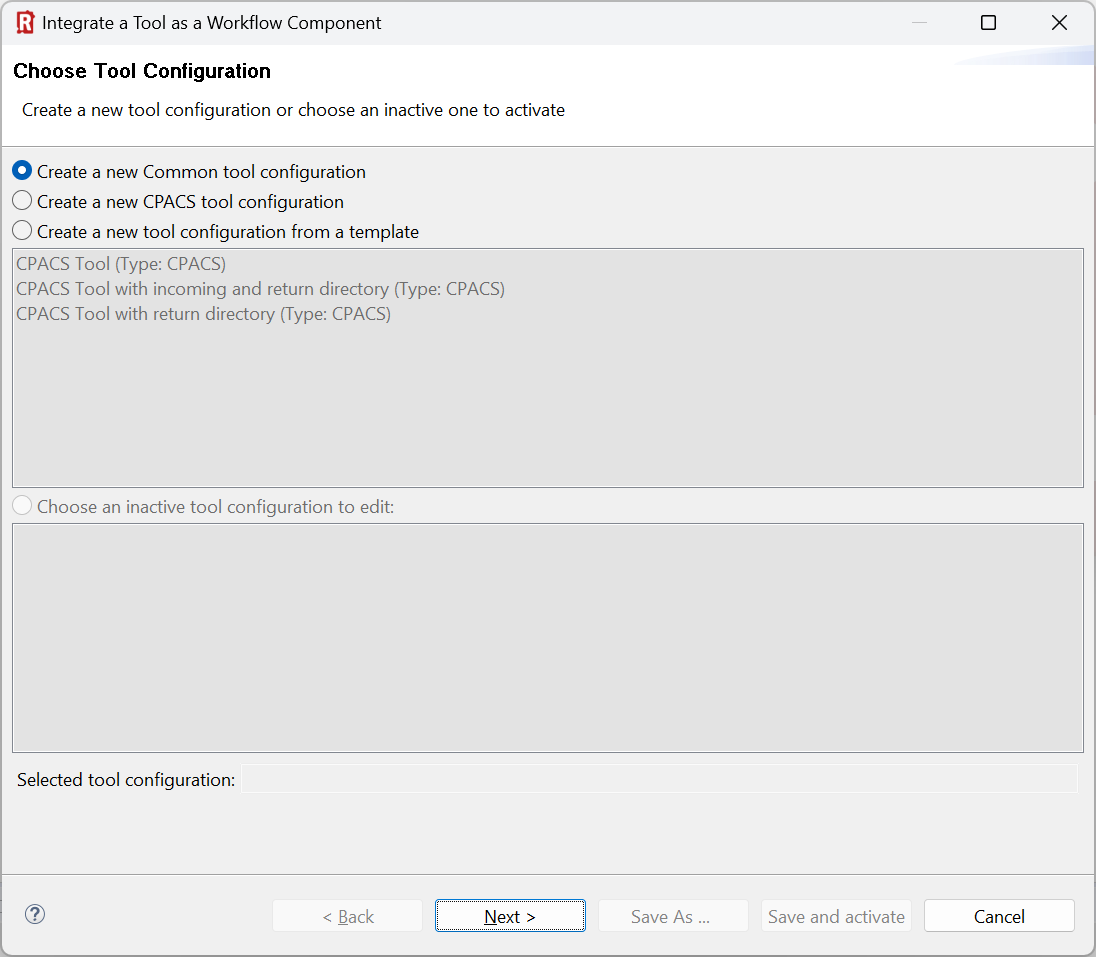}
    \label{fig:wizard:1}
  }
  \subfloat[Second Page]{
    \includegraphics[width=.32\linewidth]{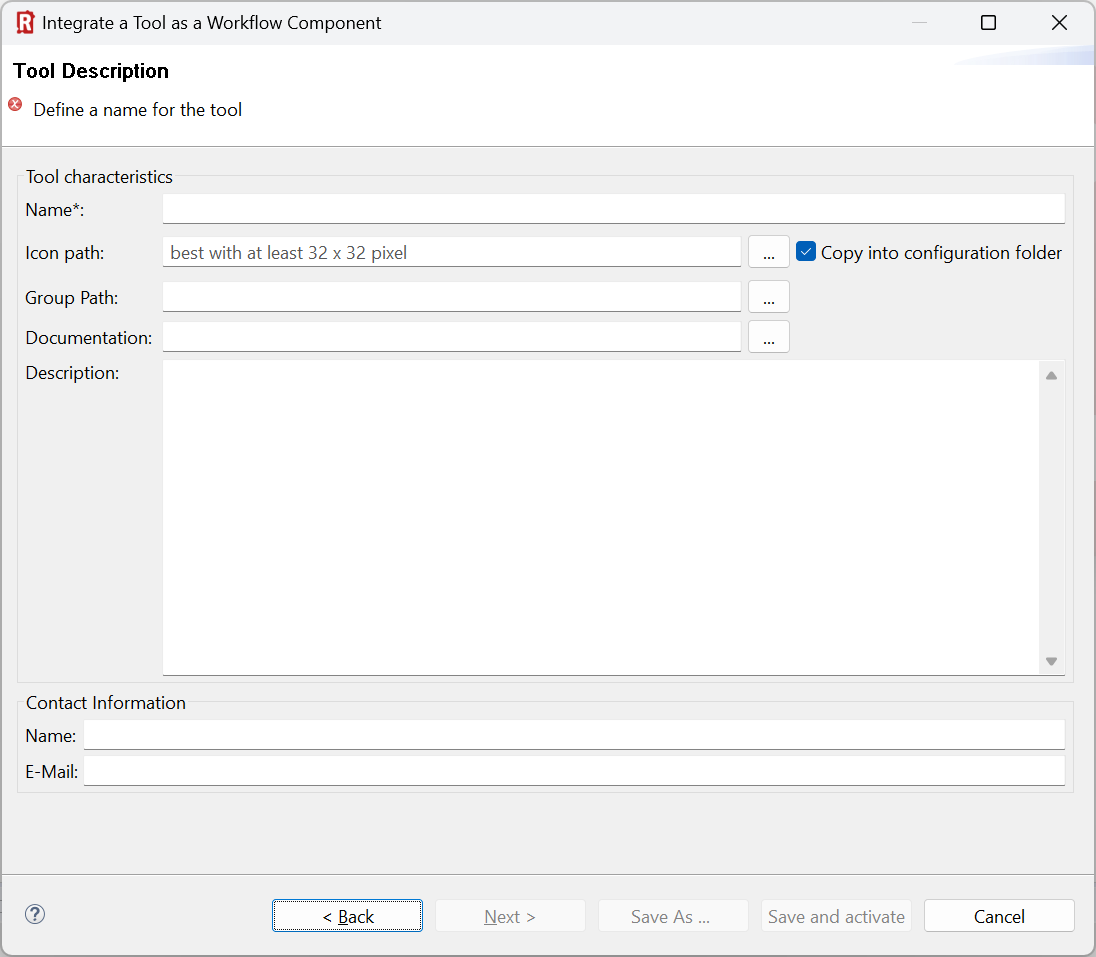}
    \label{fig:wizard:2}
  }
  \subfloat[Third Page]{
    \includegraphics[width=.32\linewidth]{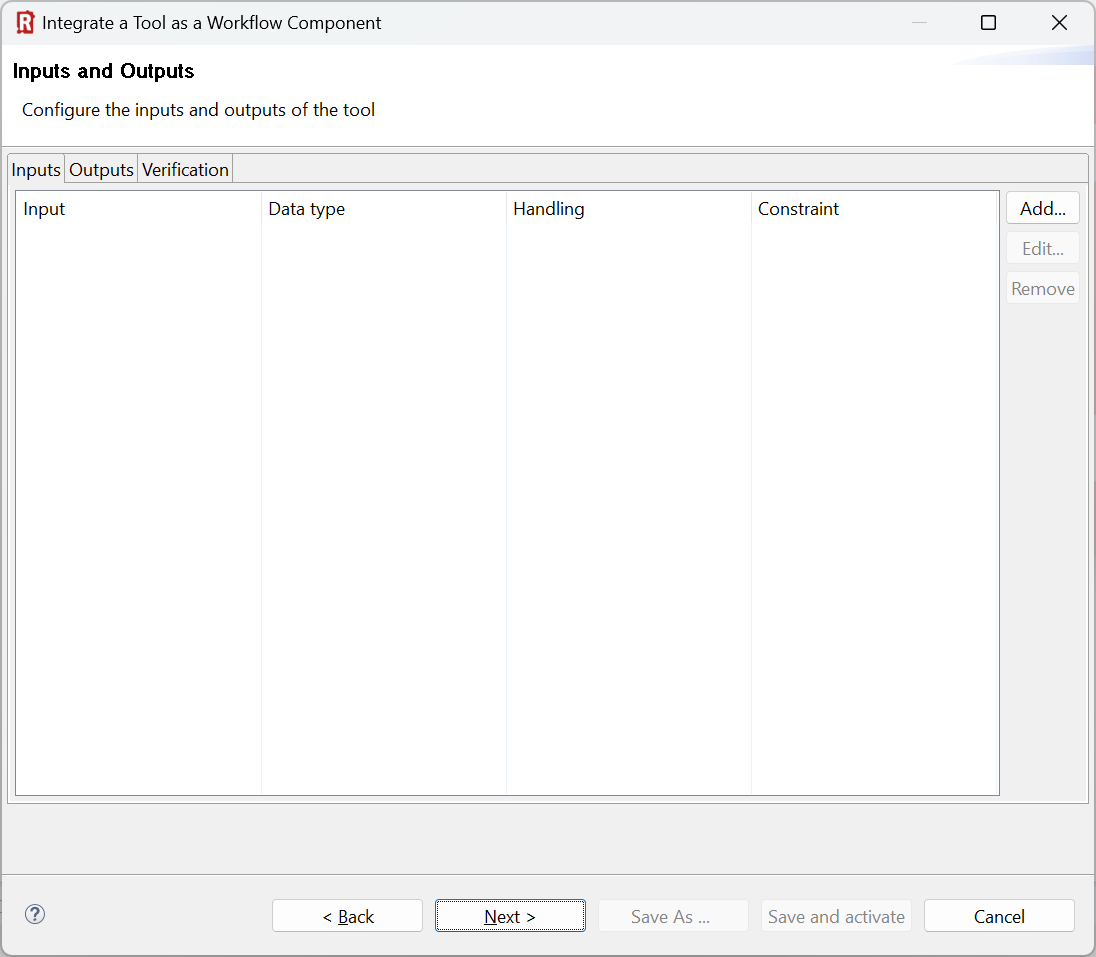}
    \label{fig:wizard:3}
  }

  \subfloat[Fourth Page]{
    \includegraphics[width=.32\linewidth]{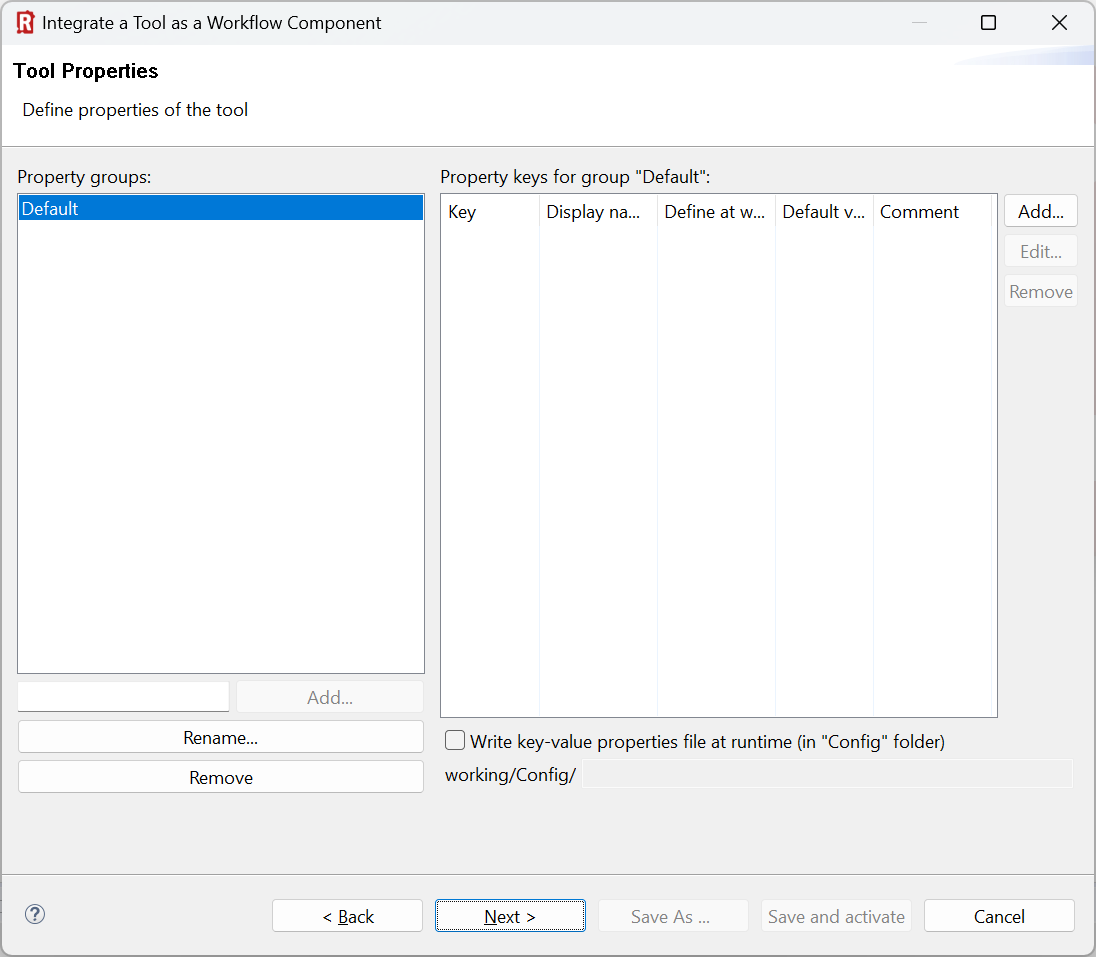}
    \label{fig:wizard:4}
  }
  \subfloat[Fifth Page]{
    \includegraphics[width=.32\linewidth]{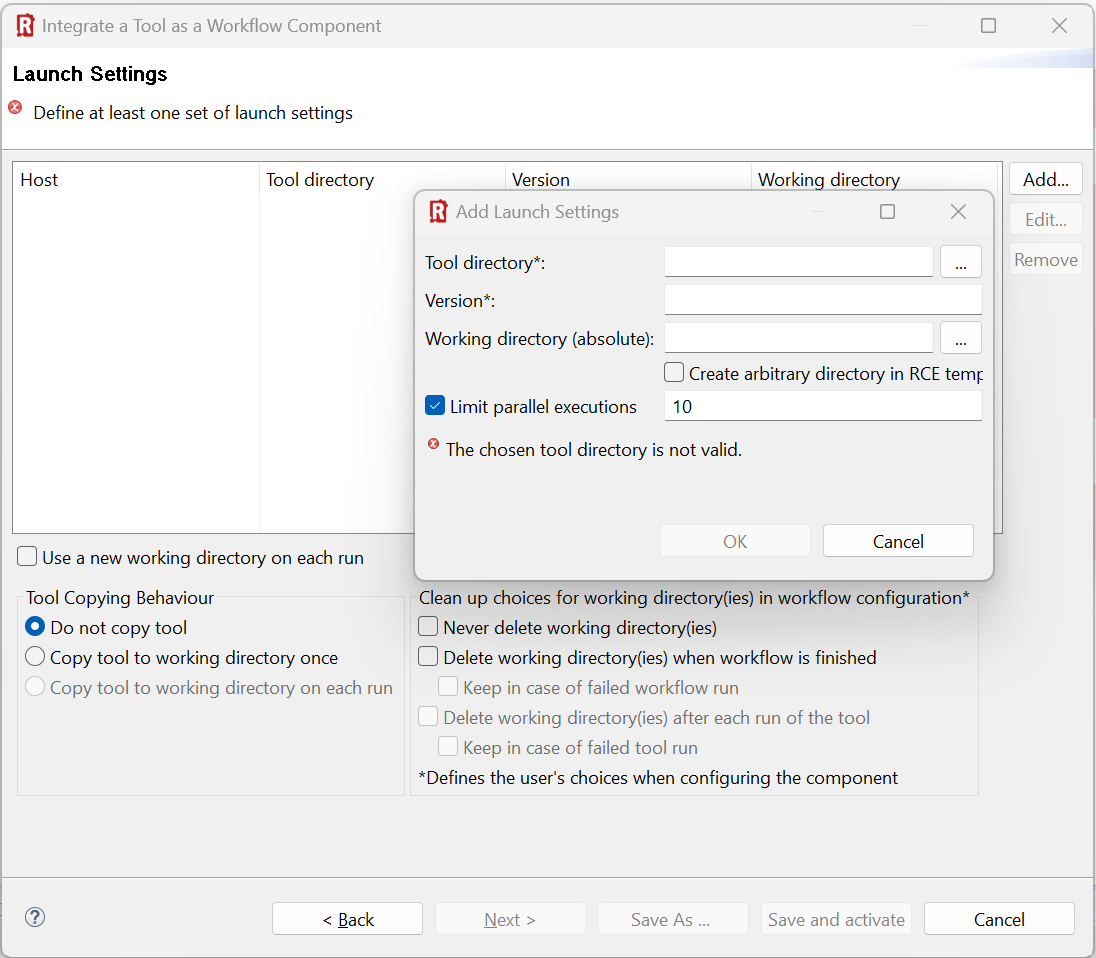}
    \label{fig:wizard:5}
  }
  \subfloat[Sixth Page]{
    \includegraphics[width=.32\linewidth]{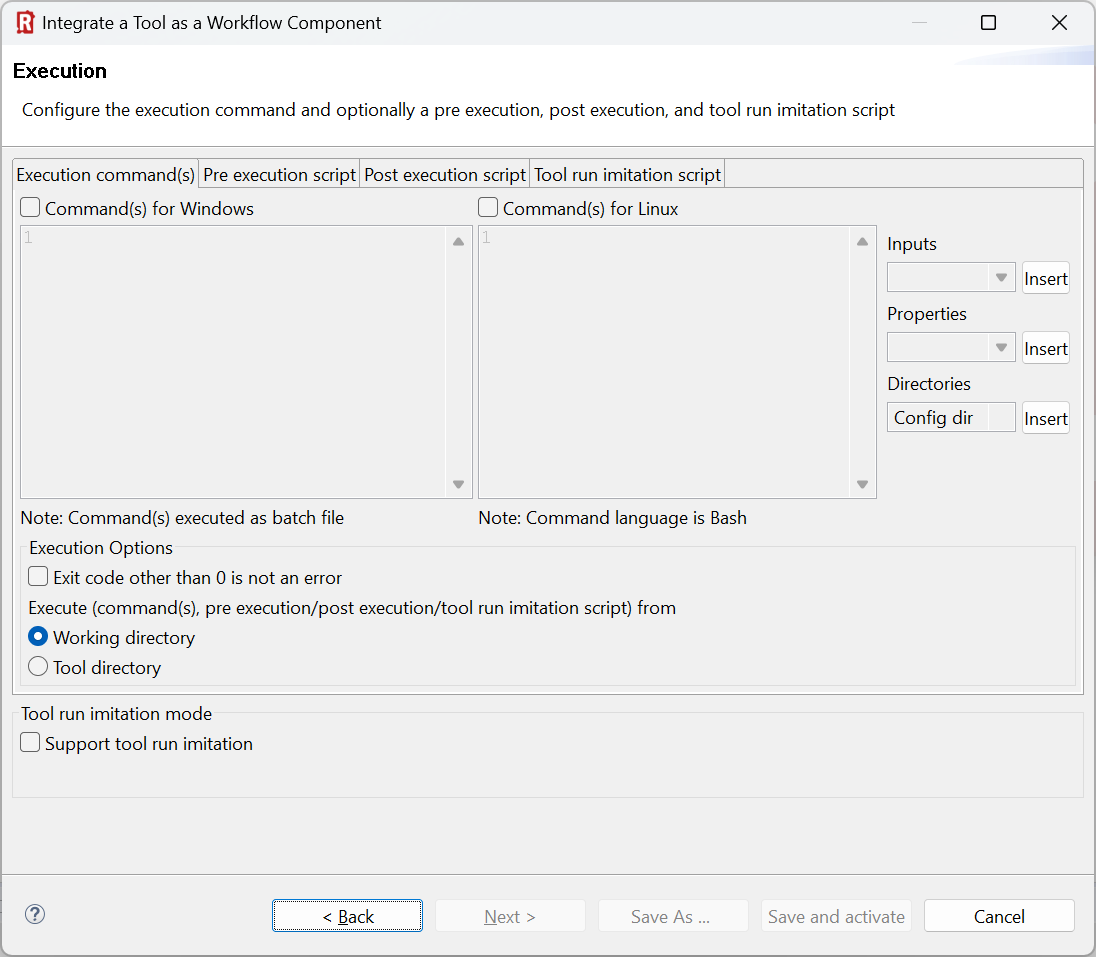}
    \label{fig:wizard:6}
  }
  \caption{Pages of the tool integration wizard of \RCE}
  \label{fig:wizard}
\end{figure*}

To simplify the integration, \RCE provides users with a desktop-based \GUI that guides the integrator through the integration.
This \GUI consists of multiple pages of \RCP-based forms which ask the user to supply the information required for integration, sometimes containing additional pop-up windows for, e.g., specifying the inputs and outputs.
We illustrate the pages of this wizard in Figure~\ref{fig:wizard}.
Keeping with the typical vernacular of desktop-based software, we call this \GUI the \emph{tool integration wizard}.

Since \RCE is used in numerous research long-running projects stability is of paramount importance.
Hence, the developers pay particular attention to quality assurance.
This takes the form not only of automated unit tests, integration tests, and end-to-end tests, but also of manual exploratory \GUI tests.~\cite{MischkeSchaffertSchneiderEtAl2022}.
The automated tests serve mainly to protect against regressions of \RCE's functionality.
In contrast, the manual tests aim to uncover unintuitive and overly complex \GUI behavior, which is hard to codify using classical testing setups.

Manual tests, while effective, require significant labor investment.
A typical interactive testing session occupies around five to seven software engineers over the course of one to three weeks, depending on the agreed-upon testing scope.~\cite{MischkeSchaffertSchneiderEtAl2022}
Hence, even partial automation of this process promises to substantially improve the development process of \RCE.
In the following section, we describe an architecture for a software tool that aims to augment the manual testing process.

\section{Testing RCE with LLMs}
\label{sec:system}

To test \RCE we have developed \GERALLT, a system focusing on using two \LLM-based agents.
This system is based on previous work by Liu \etal~\cite{LiuChenWangEtAl2024} and Gao \etal~\cite{GaoJiBaiEtAl2023}.
\GERALLT includes a \GUI Parser that converts screenshots and metadata into a structured representation by combining \OCR, icon detection, and other vision tools.
The aim of one agent is to control \RCE, while the aim of the other one is to observe the evolution of the \GUI of \RCE and to inform the user about observed inconsistencies.
After giving an overview over the architecture of \GERALLT and its constituent components in Section~\ref{sec:system:architecture} we describe the prompts used for the two \LLM-based components in Section~\ref{sec:system:controller-prompt} and in Section~\ref{sec:system:evaluator-prompt}, respectively.

\subsection{System Architecture}
\label{sec:system:architecture}

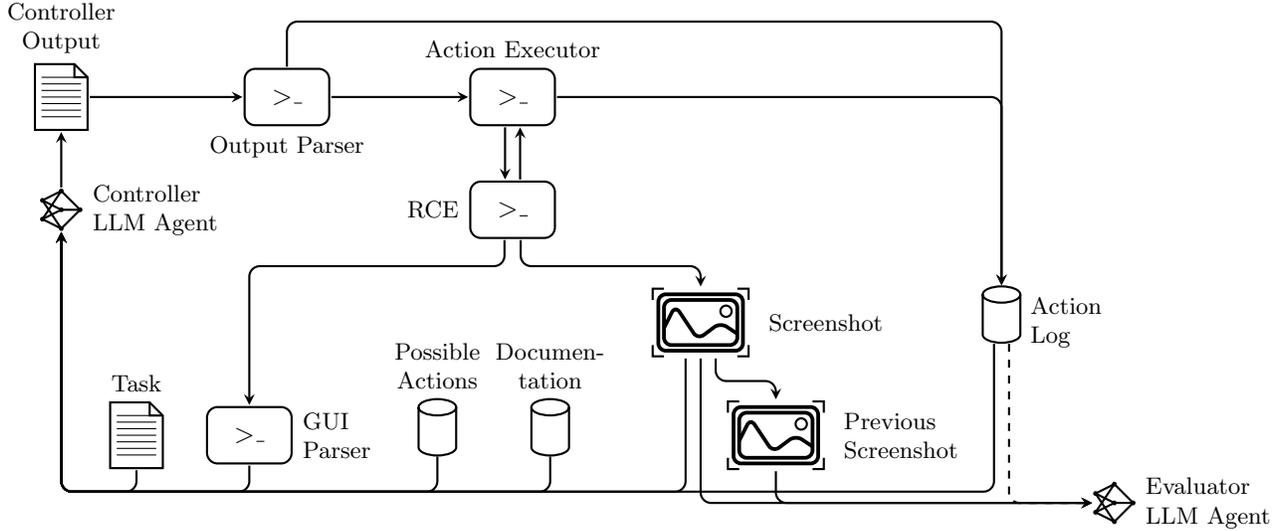
\begin{figure*}
  \centering
  \begin{tikzpicture}[thick,xscale=2,yscale=1.5,>=stealth]


  \node[document,label={[align=center]above:{Controller\\Output}}] (output) at (-3,1) {};
  \node[program,label={[align=center]below:{Output Parser}}] (output-parser) at (-1.5,1) {};
  \node[program,label={[align=center]above:{Action Executor}}] (action-exec) at (0,1) {};

  \node[program,label={[align=right]left:{RCE}}] (rce) at (0,0) {};
  \node[program,label={[align=left]right:{GUI\\Parser}}] (gui-parser) at (-1.75,-2) {};
  \node[db,label={[align=center]above:{Documen-\\tation}}] (rce-doc) at (.25,-2) {};
  \node[db,label={[align=center]above:{Possible\\Actions}}] (poss-actions) at (-.5,-2) {};
  \node[document,label={[align=center]above:{Task}}] (task) at (-2.5,-2) {};

  \node[llm,label={[align=left]right:{Controller\\LLM Agent}}] (llm-controller) at (-3,0) {};

  \node[screenshot,label={[align=left]right:{Screenshot}}] (screenshot) at (1.25,-1) {};
  \node[screenshot,label={[align=left]right:{Previous\\Screenshot}}] (screenshot-prev) at (1.75,-2) {};

  \node[db,label={[align=left]right:{Action\\Log}}] (action-log) at (3.25,-1) {};

  \node[llm,label={[align=left]right:{Evaluator\\LLM Agent}}] (llm-evaluator) at (4,-2.6) {};

  \coordinate (upper-lane) at (0,-2.5);
  \coordinate (lower-lane) at (0,-2.6);

  \begin{scope}[->,rounded corners]
    \draw (output-parser) -- (action-exec);
    \draw (output-parser.north) |- +(.5,.4) -| (action-log);
    \draw ([xshift=-.05cm]action-exec.south) -- ([xshift=-.05cm]rce.north);
    \draw ([xshift=.05cm]rce.north) -- ([xshift=.05cm]action-exec.south);
    \draw (action-exec) -| (action-log);
    \draw ([xshift=-.1cm]rce) |- +(-.5,-.5) -| (gui-parser);
    \draw ([xshift=.1cm]rce) |- +(.5,-.5) -| (screenshot);
    \draw (gui-parser) -- (gui-parser|-upper-lane) -| (llm-controller);
    \draw (rce-doc) -- (rce-doc|-upper-lane) -| (llm-controller);
    
    \coordinate (helper) at ([xshift=.1cm]screenshot.south);
    \draw (helper) |- +(.25,-.2) -| (screenshot-prev);

    \draw (poss-actions) -- (poss-actions|-upper-lane) -| (llm-controller);
    \draw (task) -- (task|-upper-lane) -| (llm-controller);
    \coordinate (helper) at ([xshift=-.1cm]screenshot.south);
    \draw (helper) -- (helper|-upper-lane) -| (llm-controller);
    \coordinate (helper) at ([xshift=-.05cm]action-log.south);
    \draw (helper) -- (helper|-upper-lane) -| (llm-controller);

    \draw (llm-controller) -- (output);
    \draw (output) -- (output-parser);

    \draw (screenshot) |- (llm-evaluator);
    \draw (screenshot-prev) |- (llm-evaluator);
    \coordinate (helper) at ([xshift=.05cm]action-log.south);
    \draw[dashed] (helper) |- (llm-evaluator);
  \end{scope}


\end{tikzpicture}
  \caption{%
    Architecture of \GERALLT. %
    The component ``Previous Screenshot'' holds the screenshot of the \GUI taken during the last iteration. %
    It is replaced by an updated screenshot after each iteration. %
    The dashed line in the bottom right denotes that the evaluator only receives the last action performed on the \GUI instead of the complete log. %
  }
  \label{fig:architecture}
\end{figure*}

We present the architecture of \GERALLT in Figure~\ref{fig:architecture}.
The aim of \GERALLT is to mimic the behavior of a human tester, i.e., to execute a loosely defined task with \RCE and to provoke unintuitive behavior while doing so.
These loosely defined task only define a goal and not concrete steps to achieve it.
Moreover, the human tester is given no guidance except for the existing documentation and their personal intuition.
We separate the two tasks of controlling \RCE and of determining unintuitive behavior into two \LLM-based components, which we call the \emph{controller} and the \emph{evaluator}, respectively.
The controller is responsible for executing meaningful actions on the GUI
The evaluator checks the GUI for issues after each action are performed.

On startup, \GERALLT takes a task as input and initializes an instance of \RCE.
In each iteration, \GERALLT constructs a prompt for the controller comprised of 
\begin{enumerate*}[label=\alph*)]
  \item the task originally given to \GERALLT,
  \item a structured description of the current state of the \GUI produced by a \GUI Parser,
  \item a list of actions possible on the \GUI widgets available in the current state,
  \item documentation given by \RCE,
  \item a screenshot of the current state of the \GUI (if the \LLM can interpret images), and
  \item the actions previously taken by the controller.
\end{enumerate*}
We describe the construction of this prompt in more detail in Section~\ref{sec:system:controller-prompt}.
\GERALLT gives this prompt to the controller, which outputs a selection of one of the possible actions.
Afterwards, \GERALLT parses the output of the controller and tries to execute the selected action on the \RCE instance.
It moreover records the attempted action as well as whether or not it was successfully executed in an action log.

After the execution was attempted, \GERALLT constructs a prompt for the evaluator comprised of
\begin{enumerate*}[label=\alph*)]
  \item a screenshot of the \GUI before the action was attempted,
  \item a screenshot of the \GUI after the action was attempted, and
  \item a description of the attempted action.
\end{enumerate*}
We describe the construction of this prompt in more detail in Section~\ref{sec:system:evaluator-prompt}.
The prompt moreover contains a request to determine unintuitive behavior of the \GUI.

\subsection{Controller Prompt}
\label{sec:system:controller-prompt}

In this section we describe the prompt given to the controller agent.
We showcase an instance of this prompt in Table~\ref{tab:controller-prompt}.
The overall structure of the prompt follows a classical pattern as described by Peckham, Jeff, et al.~\cite{PeckhamDayHbr2024}.
We first provide relevant context about the role of the LLM as well as a concrete description of its task.
Afterwards, we give examples of the desired interactions as well as a history of previous user interactions.
Finally, we conclude the prompt by concisely reiterating the task.

\begin{table*}
  \centering
  \caption{Composition of the controller prompt. Repetitions omitted and denoted by ellipses. Newlines omitted where possible.}
  \label{tab:controller-prompt}
  \scriptsize
  \begin{tabular}{lp{12cm}} \toprule
  Part & Example \\ \midrule
  Role & Your Role: 
    You are an automated system that controls the software RCE.
    You are given a task, the GUI of RCE and documentation about the software in textual form.
    You have to interact with the GUI to achieve the given task.
    You can control the GUI by sending actions to the software.
    The software will execute the actions and give you feedback about the result, by sending you the new state of the GUI and whether the action was successful or not.
    You must use the information about the GUI and the documentation to decide which actions to take.
    Include the information about the position of the GUI-Elements and the text of the GUI-Elements in your decision.
    Also consider which Parent-Elements the GUI-Elements have.
    It is important to take the context of the GUI-Elements into account when deciding which actions to take
    You can also use the feedback about the result of the actions to decide which actions to take next. \\ \midrule
  Task & Your task:
    Act as a GUI-Tester for the software RCE.
    Cover all pages of the Tool Integration Wizard.
    Use as many different UI-Elements as possible.\\ \midrule
  Documentation & Documentation:
    \# CPACS Tool Properties (optional)
    \#\# Synopsis
    Configure the CPACS tool specific values.
    \#\# Usage
    Fill in the following values to make your tool using the CPACS specific features, e.g.input and output mapping.
    - **Incoming CPACS endpoint name**: Select theinput that represents the incoming CPACS file. This input must be configuredon the "Inputs and Outputs" page.
        **Note**
        The data type of this input must be "File". Usage must be "required".
    - **Input mapping file**: Select or enter the mapping file for input mapping. Supported file extensions are ".xml" for classic mapping and ".xsl" for advanced XSLT-mapping.
        **Note**
        The path must be relative to the tool directory configured on the "Launch Settings" page. The file chooser dialog regards thatfact. [...] \\ \midrule
  GUI & The current state of the GUI:
  \{
      "class\_name": "Dialog",
      "control\_type": "WindowSpecification",
      "control\_id": 0,
      "rectangle": [ "L4429", "T1655", "R5172", "B2299" ],
      "text": "Integrate a Tool as a Workflow Component",
      "sub\_elements": [ ... ] \} \\ \midrule
  Action & There are different types of GUI-Elements in the GUI of RCE.
    UIAWrapper and StaticWrapper are GUI-Elements that can not be interacted with.
    Their only purpose is to display information or group other Gui Elements.
    The ButtonWrapper is a GUI-Element that can be clicked. ... \\ 
         & To control a GUI-Element output a command in the following Format:
              \textlangle action \textrangle (\textlangle control id\textrangle), for example click(134478)
          For each control type there are different actions posible.
          The StaticWrapper, ToolbarWrapper and UIAWrapper have no actions.
          The ButtonWrapper has the click(\textlangle control\_id\textrangle) action, for example click(134478) ... \\
         & You must format your output in JSON as the following:
          {
              "action": "\textlangle action\textrangle",
              "explanation": "\textlangle what the action does and why you do it\textrangle"
          }
          example 1:
          {
              "action": "click(134478)",
              "explanation": "click next to get to the second page"
          } ... \\ \midrule
  Action Log & The previous actions:
[
  {
      "action": "write(2166788, 'Number of Threads')",
      "explanation": "Enter a display name 'Number of Threads' for the 'numThreads' property to make the key human-readable and provide context in the properties view.",
      "status": "executed"
    }, ... ] \\ \midrule
    Closing & What action do you want to take to do the next step for achieving the given task?
  \\ \bottomrule
  \end{tabular}

\end{table*}

In the first section of the prompt, we provide context about the role of the \LLM.
For this, we describe that the \LLM will receive a concrete task and textual information about the current state of the \GUI of \RCE and that it is expected to pick to next action to perform on the \GUI in order to achieve its stated task.
Recall that we aim to have \GERALLT simulate the behavior of a new and non-expert user of \RCE.
Hence, we omit descriptions of the intended use of \RCE as well as of its functionality on purpose.

Having described the context of \GERALLT, we state its concrete task, namely to ``act as a \GUI-Tester for the software \RCE''.
We moreover provide the evaluation criterion of covering ``as many different \GUI-elements as possible''.
Again, we consciously omit more concrete definitions of steps to take.
This serves having \GERALLT emulate the manual testing, which aims to explore all possible interactions of \GUI elements.

\begin{figure}
  \includegraphics[width=\linewidth]{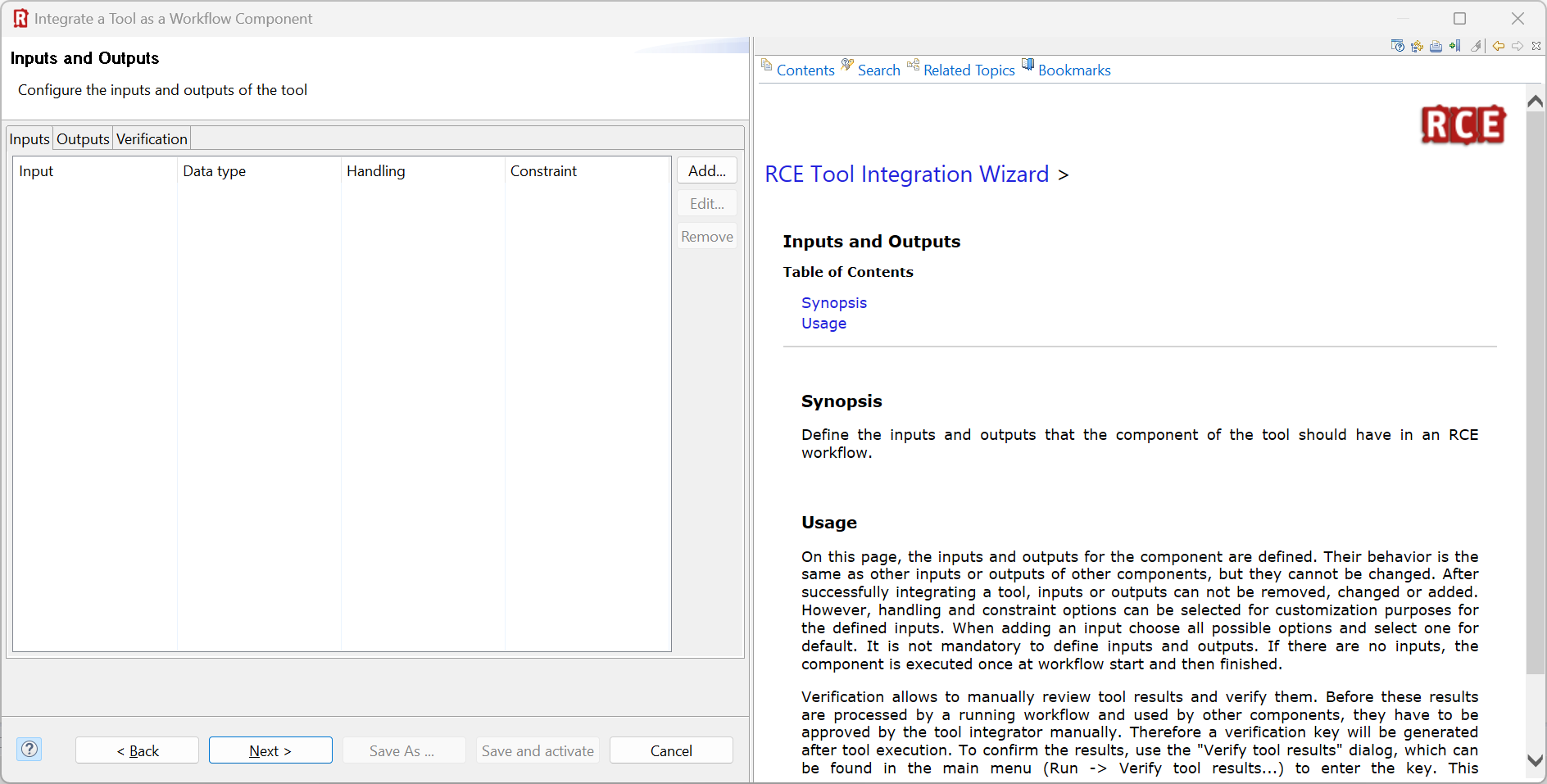}
  \caption{The appearance of the online help in the tool integration of \RCE.}
  \label{fig:online-help}
\end{figure}

Initial experiments with \LLMs showed that the \LLM needs more information about \RCE other than the \GUI information.
This manifested in the controller agent initially failing to proceed through the various stages of the tool integration.
This, in turn, was mainly due to to the text emitted by the \LLM to the \GUI not clearing input sanitization, e.g., not providing a well-defined path when requested.
To overcome this obstacle, we include relevant parts of the \RCE documentation in the prompt.
This documentation is included in \RCE and also available to human testers.
We illustrate the appearance of this online help in the right-hand side of Figure~\ref{fig:online-help}.

Afterwards, we provide the \LLM with a textual description of the current state of the \GUI.
This description is generated by the \emph{GUI Parser}.
The GUI Parser uses the PyWinAuto Python package~\cite{pywinauto2019} to extract the \GUI elements on \OS level.
This works similar to the approach of Liu \etal~\cite{LiuChenWangEtAl2024}, but instead of Android the \OS is Windows.
The description takes the form of a \JSON representation of the widgets comprising the \GUI.
As each widget can contain multiple child-widgets, this representation describes a tree of widgets.
Each widget is described by its name, its type (e.g., a static text, a combo box, or a text input field), its unique ID, its position on the screen.
Additionally, some widgets have additional information, e.g. a checkbox has the information if it is checked.

This concludes the description of the current state of the system under control.
It remains to define the actions available to the \LLM.
To this end, we define for each type of element a list of possible actions.
Moreover, we state that the output of the \LLM shall conform to a given \JSON schema to simplify subsequent parsing.
In addition, we provide a list of previously taken actions to make the evolution of the \GUI up to this point explicit.
This way, we do not have to rely on the context window of the \LLM to retain the previously taken actions.

Finally, we conclude the prompt with a concrete question about the next action to be executed on the \GUI.
This concludes the description of the prompt for the controller \LLM and the rationale behind its construction.
In the next section we describe the prompt for the evaluator \LLM.

\subsection{Evaluator Prompt}
\label{sec:system:evaluator-prompt}

The aim of the evaluator is to ``observe'' the effect of the controller's actions on the \GUI of \RCE and to determine whether any actions prescribed by the controller have an unintuitive consequence on the \GUI.
For this, the evaluator receives as input a screenshot from before the action and one after the action together with a textual prompt.
The prompt again consists of an initial section describing the context and concrete task of the component.
Similarly to the documentation of \RCE provided to the controller, the task description contains a list of common errors and unintuitive behaviors typically caught by human testers.
The prompt then contains instructions on formatting the output as well as the most recent action prescribed by the controller.
This description is taken directly from the explanation produced by the controller.

\begin{table*}
  \centering
  \caption{Composition of the evaluator prompt. Newlines omitted where possible.}
  \label{tab:evaluator-prompt}
  \scriptsize
  \begin{tabular}{lp{12cm}} \toprule
  Role  &
  You are a very experienced GUI Tester.
You observe the GUI of a Software.
A system performs actions on the GUI.
After each action taken on the GUI you evaluate whether the software behaves as expected or if there are any issues.
You are provided with a screenshot of the GUI before and after the action. \\ \midrule
  Task &
Check if there are any inconsistencies, unexpected behaviour or UI Elements that are not visible.
In detail check the following:
- the size, position, height, width of the visual elements
- Checking the message displayed, frequency and content
- Checking alignment of radio buttons, drop downs
- Verifying the title of each section and their correctness
- Cross-checking the colors and its synchronization with the theme \\ \midrule
  Output &
Format your output as JSON.
For example:
\{
    "state": "problem",
    "reason": "The Text of the Description is not fully visible."
\}
If there are no problems output:
\{
    "state": "okay"
\} \\ \midrule
Action & 
The action performed between the two images was was:
Click the 'Next \textrangle' button to proceed to the next page of the Tool Integration Wizard and continue testing the remaining pages and their functionalities. \\ \midrule
Closing & Are there any problems with the GUI after the action was performed? \\ \bottomrule
\end{tabular}

\end{table*}

Since the prompt for the evaluator follows a similar structure and analogous reasoning to the controller prompt, we omit a detailed description.
Instead, we provide an instance of this prompt in Table~\ref{tab:evaluator-prompt}.
This concludes the description of the architecture of \GERALLT.
In the following section, we evaluate \GERALLT and discuss its known limitations.

\section{Evaluation and Known Limitations}
\label{sec:evaluation}

To evaluate \GERALLT, we implemented the individual components described in Figure~\ref{fig:architecture}.\cite{GERALLT}
We implemented most components of \GERALLT as Python scripts, namely the Output Parser, the Action Executor and the construction of the controller prompt and evaluator prompt.
For the implementation of the controller agent and evaluator agent, previous work by Rosenbach~\cite{Rosenbach2024} determined ChatGPT by OpenAI~\cite{OpenAI} to be the most promising \LLM for this use case.
So we use the GPT-4o model over OpenAI's API, so that \GERALLT runs without human intervention.
Since ChatGPT is multi-modal, we are able to supply a screenshot of the current state of the \GUI to the \LLM in addition to the textual prompt, as described in Section~\ref{sec:system:controller-prompt}.
GPT's responses are very variable.
However, this is beneficial for covering more parts of the tested software application.
Moreover, we used PyWinAuto~\cite{PyWinAuto} to implement the \GUI Parser and the Action Executor.

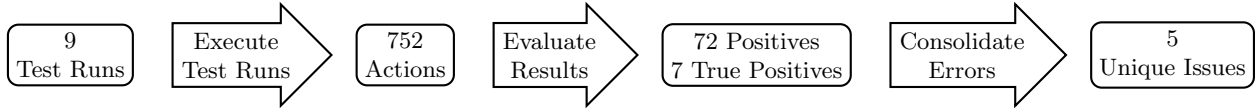
\begin{figure*}
  \centering
  \begin{tikzpicture}[thick]
  \node[draw,align=center,anchor=west,rounded corners] (test-runs) at (0,0) {9\\Test Runs};
  \node[single arrow, draw,align=center,anchor=west] (execute-runs) at ([xshift=.5cm]test-runs.east) {Execute \\ Test Runs};
  \node[draw,align=center,anchor=west,rounded corners] (actions) at ([xshift=.3cm]execute-runs.east) {752\\Actions};
  \node[single arrow, draw,align=center,anchor=west] (evaluate) at ([xshift=.5cm]actions.east) {Evaluate\\Results};
  \node[draw,align=center,anchor=west,rounded corners] (positives) at ([xshift=.3cm]evaluate.east) {72 Positives \\ 7 True Positives};
  \node[single arrow, draw,align=center,anchor=west] (consolidate) at ([xshift=.5cm]positives.east) {Consolidate\\Errors};
  \node[draw,align=center,anchor=west,rounded corners] (issues) at ([xshift=.3cm]consolidate.east) {5\\Unique Issues};
\end{tikzpicture}
  \caption{The evaluation process for our system.}
  \label{fig:evaluation}
\end{figure*}

Recall that it was our aim to construct a tool that supports software developers in assuring the quality of \RCE.
In particular, we aimed to support developers in performing time-intensive exploratory \GUI tests.
These tests are targeted at finding ``unintuitive'' or ``unnatural'' behavior of the \GUI of \RCE as well as finding functional errors.
Hence, there exists no baseline of existing or known errors against we can evaluate the results of our tool.
Instead, we performed a qualitative evaluation of our tool and used the judgment of the software developers as an alternative to an objective ground truth.
We illustrate the complete evaluation process in Figure~\ref{fig:evaluation}.

\subsection{Quantitative Evaluation Results}

We conducted nine isolated test runs.
Each test run consisted of a fresh instance of \GERALLT.
Each of these instances was given the same initial prompt as described in Section~\ref{sec:system:controller-prompt} and Table~\ref{tab:controller-prompt}.
Since the \LLM-based components of \GERALLT are inherently non-deterministic, each test run resulted in different judgments given by the evaluator component.

During these test runs, the controller performed 752 actions on \RCE.
After each action, the evaluator determined whether it deemed the effect these actions had on the \GUI as problematic as described in Section~\ref{sec:system:evaluator-prompt} and in Table~\ref{tab:evaluator-prompt}.
The output of the evaluator stated a problematic result after 72 actions.

We then manually separated the outputs of the evaluator into true and false positives, i.e., into those that correctly describe the current state of the \GUI and those that do not.
This separation process resulted in seven true positive issues.
We moreover determined two pairs of issues to result from the same underlying cause.
Thus, the evaluation process results in five unique issues with the tool integration wizard of \RCE.
The problems range from optical problems to functional errors.
We discussed the five issues with the development team of \RCE who confirmed that they agreed with the classification given by our tool.
Moreover, the developers agreed that these issues constituted ``blind spots'' in the existing testing process, i.e., that they did not consciously notice these issues during manual testing.

\subsection{Example Issue}

\begin{figure*}
  \subfloat[Screenshot before the performed action]{
    \includegraphics[width=0.45\linewidth]{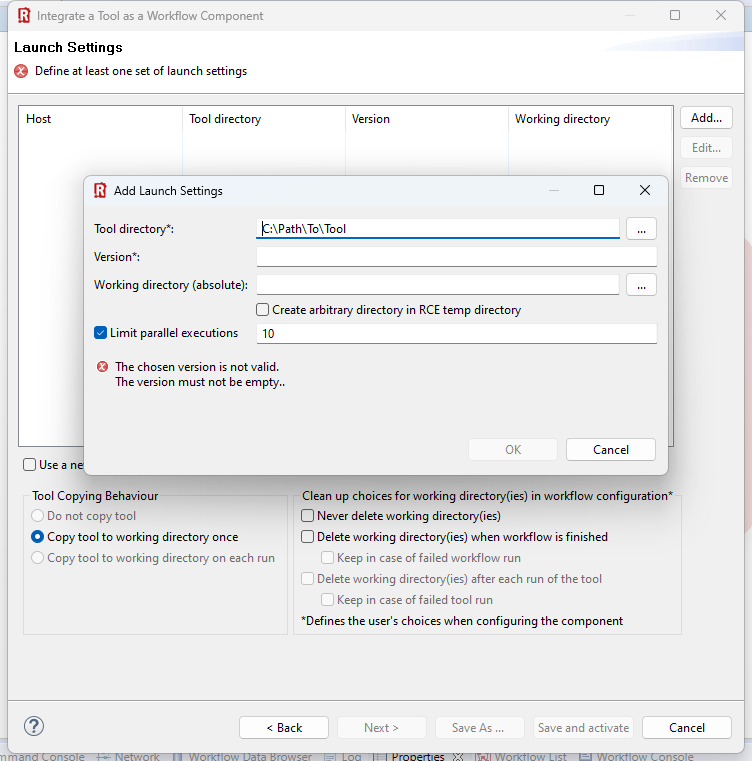}
    \label{fig:screenshot_before}
  }
  \subfloat[Screenshot after the performed action]{
    \includegraphics[width=0.45\linewidth]{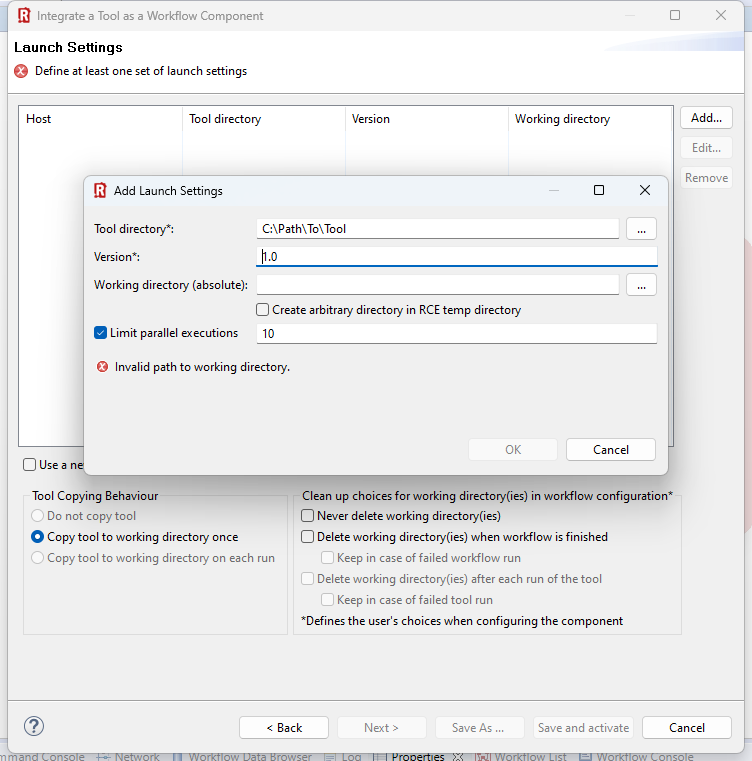}
    \label{fig:screenshot_after}
  }
  \caption{Example error, where the evaluator criticized that the error message is too imprecise.}
  \label{fig:example_problem}
\end{figure*}

We show an example problem found by the evaluator in Figure~\ref{fig:example_problem}.
Here, the controller has opened the pop-up window asking the tool integrator to provide the \emph{launch settings} of the tool to be integrated.
These launch setting comprise 
\begin{enumerate*}[label=\alph*)]
  \item the directory in which the tool is installed,
  \item the version of the tool,
  \item the absolute path to the working directory in which the tool shall be executed, and
  \item the maximal number of instances of the tool that can be executed in parallel.
\end{enumerate*}
The controller has moreover entered a path to the tool directory.
The validation of the contents of the pop-up window proceeds from top to bottom.
Hence, \RCE informs the user about a problem with the empty field for the tool version, namely that ``The chosen version is not valid. The version must not be empty..[sic]''

During the next iteration of \GERALLT, the controller inputs the string ``1.0'' into the field ``Version''.
Hence, \RCE determines the supplied version to be valid.
Since input validation proceeds with the next field in the form, \RCE now informs the user that the (non-existent) path to the working directory is invalid.

The evaluator agent determines that this behavior is problematic and returns the explanation ``The error message 'Invalid path to working directory' is visible even though no path was provided, which is inconsistent unless a path was entered.''
In other terms, the evaluator criticizes that a non-existent path cannot be invalid.
Moreover, it criticizes that in the case of an empty version field, the user was informed about the field being empty (cf. Figure~\ref{fig:screenshot_before}), while they are given no such specific error message in the case of an empty working directory field (cf. Figure~\ref{fig:screenshot_after}).

\subsection{Known Limitations}
\label{sec:evaluation:limitations}

In the previous sections, we have presented the capabilities of \GERALLT.
We now turn our attention to discussing its limitations, particularly in terms of the system requirements of the implementation, the system under test, and the software qualities tested by \GERALLT.

The current implementation of our system is limited to testing \RCE on Windows.
This is due to the choice of PyWinAuto~\cite{PyWinAuto} for the implementation of the \GUI parser, which only allows automation of the Windows \GUI.
GNOME- or KDE-based \GUIs could be tested using, e.g., Dogtail~\cite{Dogtail} or Appium~\cite{Selenium,Appium}.
This would require some implementation effort, but no change to the overall system architecture presented in this work.

In this work, we use \GERALLT to test the tool integration wizard of \RCE.
This requires the user to perform left clicks and text inputs.
Other features of \RCE require more advanced interactions such as, e.g., dragging and dropping interface elements.
Similarly to the previous case, implementing such interactions would require adaptations to the Output Parser as well as the Action Executor.
However, no conceptual change of the architecture presented in Figure~\ref{fig:architecture} would be required.

Finally, \GERALLT only aims to find \GUI behavior that human users would deem unintuitive and functional errors.
In particular, we have not constructed the system to determine, e.g., accessibility or security issues exhibited by the system under test.
Detecting such issues is out of the scope of this work.
Moreover, in our opinion, such measures of quality are better tested for using rule-based, deterministic approaches.

\section{Conclusion and Future Work}
\label{sec:conclusion}

In this work we have given preliminary evidence that an \LLM-based automated testing system can support software developers in exploratory \GUI testing.
Our developed system called \GERALLT aims to determine behavior of the \GUI that is deemed unintuitive or surprising by software developers.
Moreover, the issues identified by \GERALLT can find issues in ``blind spots'' of the software developers.
Our work thus provides an interesting new direction for circumventing such blind spots in exploratory testing.

Our work constitutes a promising contribution to the state of the art.
There are numerous avenues in which future research can continue.
Our main aim in these future research directions is to evaluate \GERALLT against existing manual testing processes.
This requires objective measures of the efficiency and effectiveness of exploratory testing.

Moreover, we are looking to increase the applicability of \GERALLT.
For this, we are following three major directions: Making the system applicable to other features of \RCE, using it on the same software running on additional operating systems, and testing it on other engineering software products.
As discussed in Section~\ref{sec:evaluation:limitations}, these extensions mainly amount to additional work on the implementation instead of significant changes to the architecture presented in Section~\ref{sec:system:architecture}.

We believe that our contribution in this work serves as an important stepping stone towards the development of \LLM-based testing systems.
These systems will provide valuable feedback on the intuitiveness of the \GUI to developers without laboriously involving human testers.
Such systems will not replace human feedback, but instead augment the testing process.
Eventually, the aim of developing these systems is to move acceptance testing ever further ``left'' in the development cycle, thus decreasing the cost of software development and increasing software quality.

\balance

\bibliographystyle{IEEEtran}
\bibliography{2025-05-ast-llm-rce-integration}

\end{document}